# Spins in Semiconductor Nanocrystals

Gustavo Martini Dalpian

Center for Natural and Human Sciences,

Federal University of ABC,

Santo André, SP, Brazil

email: gustavo.dalpian@ufabc.edu.br

**Abstract**

*Semiconductor nanocrystals are being used as hosts to trap and manipulate single spins. Spins in nanocrystals can have different properties than their bulk counterparts, owing both to quantum confinement and surface effects. We will show that spins can be generated in nanocrystals by the insertion of impurities, either magnetic or not, or by magnetic polarons present at dangling bonds at their surfaces. This chapter reports theoretical contributions to this field, where the Density Functional Theory is used to simulate these functionalized nanocrystals.*

## 1. Introduction

During the last 30 years, the development of nanoscience lead to the possibility of manipulating nanostructures at the atomic level. Nanocrystals can be grown in a wide variety of geometries and shapes, and for many of the known compounds, leading to the possibility of fine-tuning their properties. When the *relatively-new* field of nanoscience met with the mature area of magnetism, new phenomena arose. Having magnetic spins in nanocrystals leads to the possibility of controlling and manipulating single spins in an atomic level. Besides the exciting fundamental challenges related to single-spin manipulation, there is also a wide variety of possible applications of this type of technology, going from spin filters (Efros et al 2001) to single electron transistors (Qu et al 2006). The ability to manipulate quantum processes at the level of single impurities forms the basis of an area of research called solotronics (Kobak et al 2014).

Understanding the nature of spins in semiconductor nanocrystals is, consequently, an important task in order to successfully manipulate them. Theoretical and computational tools based on the Density Functional Theory (DFT) and on pseudopotentials methods are a suitable choice for modeling these materials since they combine precision with the possibility of simulating the nanostructures in similar ways as they occur in the laboratory. Modern methodologies can simulate nanostructures with thousands of atoms, both with and without periodic boundary conditions.

In this text, we will qualify two different types of spins in semiconductor nanocrystals and report theoretical, DFT-based results on them. The first type of spin involves the insertion of extrinsic elements into the semiconductor host, in a procedure very similar to what is done for dilute magnetic semiconductors (Dietl 2010). The most popular impurities are transition metal atoms, which usually have large exchange splittings, leading to large local magnetic moments. Besides these magnetic elements, it is also possible to observe spins and a magnetic response by inserting non-magnetic impurities into the nanocrystal, such as carbon or nitrogen in ZnO. This might have advantages since standard elements used for magnetic semiconductors are often toxic and very expensive.

The second way of obtaining spins in nanocrystals is through the intrinsic existence of surfaces in these materials. Dangling bonds present at these surfaces might act as localized spins of magnetic ions, producing a similar effect as the insertion of transition metals into the host nanocrystal.

## 2. Doped Nanocrystals

Most bulk semiconductors can also be grown as low dimensional nanostructures. Even materials that have only been studied more deeply in the last few years, as the Halide Perovskites, have already been synthesized in the form of nanocrystals (Castañeda et al 2016). The same is true for other, more traditional, semiconductors such as Si, ZnO, CdTe, CdSe and so on. In order to broaden the possibility of applications of these nanostructures, it is important to functionalize them. Doping is one of the most conventional ways of functionalizing semiconductors, and should also be a suitable path for semiconductor nanocrystals. However, this can be a formidable task owing to kinetic limitations and self purification mechanisms.

Turnbull, in 1950, was the first one to argue that small crystals would have a smaller concentration of defects when compared to bulk (Turnbull 1950). He argued that defects are easily annealed out owing to the material's limited size: the defect/impurity does not have to travel too much to reach the surface of the nanocrystal. In 2005, Erwin et al (Erwin et al 2005) argued that the limiting factor for doping a semiconductor nanocrystal was the binding energy of the impurity to the surface of the nanocrystal. In 2006, Dalpian and Chelikowsky (Dalpian et al 2006) showed that the formation energy of the impurity increases as the size of the nanocrystal decreases. This is a direct consequence of the fact that the impurity levels get deeper as the size of the nanocrystal decreases. This *pinning* of the impurity levels has also been observed experimentally (Norberg et al 2006). Using DFT calculations, effective mass models and magnetic circular dichroism spectroscopy, it has been shown that the donor binding energies of $Co^{2+}$ impurities in ZnSe quantum dots are pinned irrespective of the size of the nanocrystal. This occurs owing to the smaller effect of quantum confinement in the localized $Co_d$ levels when compared to the HOMO (highest occupied molecular orbital) and the LUMO (lowest unoccupied molecular orbital) of the nanocrystal.

A lot of advances in the area of doped nanocrystals happened in the last decade, leading to the successful control of doped nanocrystals from an experimental perspective (Mocatta et al 2011).

From a theoretical perspective, in order to simulate semiconductor nanocrystals, one has to build molecular models that resemble what happens in these nanostructures. In this chapter, nanocrystals are generated by cutting spherical regions from a bulk crystal, as schematically shown in Fig 1a. This nanocrystal will be defined by its center (that can be a bond, interstitial position or an atom) and its radius. Depending on the definition of the center of the nanocrystal, its symmetry might be changed (Dalpian et al 2006). The generated structure, shown in Fig 1b, will present a high density of surface dangling bonds, which will certainly interfere in the final results. Later on this chapter, the importance of these surface dangling bonds will be discussed. However, at this point, we are mostly interested in quantum confinement effects, and we would like to avoid surface states. To do that, the nanocrystal's surface atoms are saturated with pseudo-hydrogen atoms, carefully generated with fractional charges in order to maintain electron counting. Details about this theoretical procedure were thoroughly described by Huang et al (Huang et al 2005).

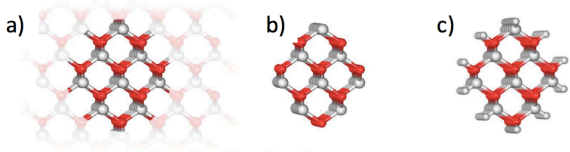

Fig. 1. Procedure used to generate spherical ZnO nanocrystals: a) and b) a spherical region is cut from the bulk crystal; c) the surface dangling bonds are saturated with pseudo-hydrogen atoms. Red and grey circles represent oxygen and Zn atoms respectively. The small, white circles represent the pseudo-hydrogen saturation.

Within the methodology described above, one can easily obtain information about the quantum confinement effects in nanocrystals, such as the increase of the bandgap as the size of the nanocrystals decreases. Quantum confinement effects on doped nanocrystals can also be explored through these methods by inserting impurities into them. This procedure will be described in the next section.

## 2.1 Magnetic impurities

To simulate magnetic impurities inside nanocrystals, we substitute one of the internal atoms of the nanocrystal by a transition metal impurity. This process has been successfully done experimentally for a variety of materials and dopants, providing a large quantity of experimental results for comparison. These results include Mn in ZnS (Donegá et al 2002), Mn in CdS (Nag et al 2010), Co and Ni in ZnO (Schwartz et al 2003), Co in ZnSe (Norberg et al 2006) and many others.

From a computational point of view, one of the first works dedicated to understand the spintronic properties of nanocrystals was performed by Huang, Makmal, Chelikowsky and Kronik in 2005 (Huang et al 2005). They studied the magnetic properties of Mn-doped Ge, GaAs and ZnSe nanocrystals using real space *ab initio* pseudopotential methodologies (Chelikowsky et al 1994), based on the density functional theory. These calculations were performed with the PARSEC code (Kronik et al 2006). In this chapter we will not detail the methodology used for performing these calculations, since this has already been discussed in other chapters. However, it is important to mention that real space methods have the advantage of avoiding the need to use large supercells, with large vacuum layers, that are usually needed in plane wave codes. The important parameter in this case is the grid spacing, which might influence in the final calculated properties (Chelikowsky et al 2011).

When Mn is inserted into these semiconductor nanocrystals, the half-metallicity trends observed in the bulk are preserved. Substituting a Ge atom by Mn will introduce $Mn_d$ orbitals into the band gap of the nanocrystal. Owing to the crystal field, these *d* levels will split into a tri-degenerated $t_2$ and a two-degenerated *e* level. The picture is very similar for GaAs and ZnSe. For Ge, there will be two holes in the $t_2$ level, whereas substituting Ga by Mn in GaAs will insert one hole. No holes are inserted when Mn substitutes Zn in ZnSe. This effect has been studied for four different sizes of nanocrystals: $X_9MnY_{10}$, $X_{18}MnY_{19}$, $X_{40}MnY_{41}$ and $X_{64}MnY_{65}$, where X=Ge, As or Zn and Y=Ge, As or Se, respectively, and the index indicates the number of atoms of that type in the nanocrystal. It was observed that, as the impurity states are highly localized, they are less affected by quantum confinement than the delocalized host states, such as the HOMO and the LUMO. As the size of the nanocrystals decrease, the Mn-related levels become deeper inside the band gap. This causes the ferromagnetic stabilization to be dominated by double exchange mechanisms via localized holes.

The insertion of Mn atoms into II-VI semiconductor does not insert carriers (holes) into the system, since Mn is isovalent to the type-II atom. Consequently, Mn substitution into ZnO nanocrystals does not induce a ferromagnetic response. However, May et al (May et al 2012) has shown that a ferromagnetic response can be obtained by the insertion of external p-type dopants. This can be done by photoexcitation of electrons or by atomic substitution. May suggested that doping ZnO quantum dots with N (at the Oxygen site) would lead to a ferromagnetic response in Mn-doped ZnO.

Owing to the localized nature of transition metal impurities, a frequent concern relies on the appropriateness of density functionals to simulate such materials. Whereas the generalized gradient approximation (GGA) or local density approximation (LDA) are very popular amongst chemists and physicists, they should not be applied in a straightforward manner for all systems. In order to elucidate this issue, Badaeva et al (Badaeva et al 2008) tested several density functionals to better understand the properties of Co-doped ZnO nanocrystals. They have compared experimental data for the dopant-carrier magnetic exchange interactions with LSDA, GGA and hybrid PBE1 functionals (Perdew et al 1997). Hybrid functionals include a fraction of Hartree-Fock exchange, leading to a better description of localized states and of bandgaps. The overall conclusion was that Hybrid functionals are usually better suited to study this type of system, although qualitative responses are already obtained with other functionals.

## 2.2  *Non-Magnetic impurities*

In the previous section we discussed spins in semiconductor nanocrystals that were created by the insertion of transition metal atoms. However, it is also possible to get spins in semiconductor nanocrystals by inserting non-magnetic atoms into the nanocrystals or by the presence of other types of defects.

One of the first theoretical reports on ferromagnetism in non-doped semiconductors was made by Pemmaraju and Sanvito (Pemmaraju and Sanvito 2005), in a study about intrinsic point defects in $HfO_2$. They have observed that Oxygen vacancies are spin-polarized in this kind of system, and that there are exchange interactions between neighboring vacancies, leading to the possibility of a macroscopic magnetization of the whole crystal. This result was important to help understand experimental observations of magnetism in non-doped samples of $HfO_2$.

As intrinsic defects, which happen spontaneously in the crystal, can lead to a macroscopic magnetization, one could use similar principles to engineer magnetism into nanocrystals without the need of transition metal atoms. This task can also be obtained by inserting non-magnetic impurities, such as Carbon or Nitrogen, into quantum dots. To test this idea, Kwak et al (Kwak et al 2009) studied Carbon-doped ZnO nanocrystals. In this case, Carbon atoms enter in substitution to oxygen, inserting two holes into the system. The ground state of C-doped ZnO is spin-polarized, with a magnetic moment of $2\mu_B$, originated from its *2p* states. As these are mainly localized states, it has been shown that quantum confinement has limited effect on the defect states, making ferromagnetism robust through different sizes of nanocrystals.

## 3.  Surface Magnetization in Semiconductor Nanocrystals

The surface saturation described previously in Figure 1c is a theoretical artifact to remove the nanocrystals' surface states from the energy gap. This is only performed to isolate the effects of quantum confinement. Although the effects of quantum confinement are modeled correctly within this model, the surface does not represent what happens in experiment. This type of saturation models a perfect saturation, where the density of dangling bonds at the surface tends to zero. Although this methodology is precise to study quantum confinement effects, we loose the rich nature of the surface of the nanocrystals. In order to analyze the surface effects on nanoparticles and their implications on properties, we have removed all surface saturation and then optimize these structures, allowing all atoms to move to their minimum energy positions. The surface of a nanocrystal can reconstruct in several new arrangements. Owing to these structural changes, and the presence of dangling bonds, the electronic structure of the nanocrystal is also altered, and a magnetic signal, or spin polarization, can also be observed. This magnetization is usually termed as $d_0$ ferromagnetism, similar to the case of non-magnetic doping described previously.

From an experimental perspective, there are several observations of ferromagnetic responses in materials that should be non-magnetic. This *phantom ferromagnetism* (Coey et al 2008) was first reported in 2008. For nanocrystals, Garcia (Garcia et al 2007) showed that non-doped ZnO nanoparticles could have a small ferromagnetic response, depending on the way the nanocrystals are saturated. When the nanocrystal is covered with Thiol molecules, a larger magnetization is observed than when the nanocrystals are covered with TOPO. These different signals are explained by different surface saturations, leading consequently to a different density of surface dangling bonds.

The nanocrystals observed in these experiments represent a saturation that is in between the two saturation limits mentioned above: i) full pseudo-hydrogen saturation and ii) no saturation. By studying non saturated nanocrystals, we will be able to learn about the properties of these dangling bonds, and infer about what happens in experiment.

Spins on dangling bonds at the surface of CdSe colloidal nanocrystals have been recently reported and studied (Biadala et al 2017). These spins have been called 'magnetic polarons'.

### 3.1 Surface reconstructions

To simulate the surface of nanocrystals, after carving a spherical nanocrystal in a bulk region, as shown in Fig 1a, we minimize the forces on all atoms letting the surfaces reconstruct. We have analyzed ZnO nanocrystals in the zinc blende (ZB) and the wurtzite (WZ) structures, and were able to observe several different types of reconstructions, and several different structural motifs for ZnO nanocrystals. We have studied different sizes of nanocrystals, with diameters ranging from 0.9nm to 1.5nm. The surface of each nanocrystal was reconstructed in a different way, giving us a broad range of different defects, including planar faces, steps, kinks, adislands, adatoms, dimers and others. We have considered the majority of possible motifs present at any nanocrystal surface.

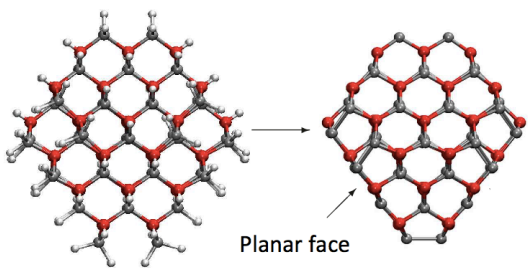

Fig. 2. Illustration of planar faces that occur in some nanocrystals after the passivation is removed. In this figure, the ZnO nanocrystal has 147 Atoms.

The most frequent motifs are the formation of planar faces and of dimers. Figure 2 shows a structural model of a saturated ZnO nanocrystal with 147 atoms, and of the non-saturated, optimized, counterpart. This planar reconstruction on ZnO has also been reported for thin films (Tushe et al 2007). Most of these reconstructions occur to reduce the dangling bond density at the surface. The formation of dimers, as also observed in Fig 2, is very common to several surfaces, including Si(001).

Besides surface reconstructions, removal of bond saturation leads to the striking observation of surface magnetization in these structures. The magnetization is strongly localized at the surfaces of the nanocrystals, with almost no signal at their interior. Figure 3 is an illustration of this phenomenon. The figure shows the radial distribution function of the spin density for different sizes of ZnO nanocrystals. The insets show the spin charge density, i.e., the total charge density for spin up minus total charge density for spin down. This radial distribution clearly shows that the magnetization in these nanostructures is mostly localized at the surface.

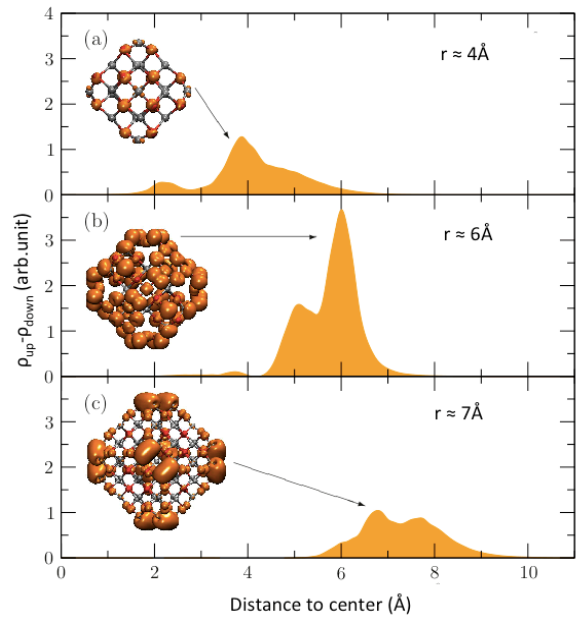

Fig. 3. Radial distribution of the spin density for three sizes of ZnO nanocrystals: a) 4 Å; b) 6 Å; c) 7 Å. The inset shows a real space representation of the spin density distribution within the nanocrystal.

Although many motifs that present magnetization were observed, the surface magnetization was not observed in all of our simulated nanocrystals. Our conclusion is that some kind of atomic order inside the nanocrystal is needed for the magnetization to be observed. For nanocrystals which structure was strongly reorganized, almost resembling an amorphous structure, no magnetization was observed (Schoenhalz et al 2009 and 2010). Besides testing different sizes of nanocrystals (with up to 238 atoms), we have also tested different crystal structures. The ground state crystal structure of bulk ZnO is the wurtzite structure, although the energy difference to the zinc blende structure is not large. On Table I we report the total magnetization of the studied nanostructures.

Table 1: Total magnetization for non-saturated ZnO nanocrystals (in Bohr magnetons). The number of atoms in the nanocrystal and the structure (ZB=zinc blende and WZ=wurtzite) is also shown. (Schoenhalz et al 2009).

| Number of atoms | Structure | Total Magnetization |
|---|---|---|
| 35 | ZB | $2\mu_B$ |
| 87 | ZB | $4\mu_B$ |
| 147 | ZB | $2\mu_B$ |
| 38 | ZB | 0 |
| 86 | ZB | $2\mu_B$ |
| 238 | ZB | 0 |
| 39 | WZ | 0 |
| 92 | WZ | $2\mu_B$ |
| 34 | WZ | 0 |
| 88 | WZ | 0 |

Other nanostructured ZnO materials have also been shown to present a ferromagnetic ordering without the need of magnetic impurities. This is the case of ZnO nanowires (Podila et al 2010), where the surface magnetization can also be used to explain the observed results.

As surfaces are delocalized and spread over the whole material, this effect should suffice to explain the observed magnetization in non-doped ZnO nanostructures (Garcia et al 2007). Other extended defects such as grain boundaries and dislocations should also be able to hold a macroscopic magnetization.

### 3.2 Interaction between surfaces and magnetic impurities

As demonstrated so far, spins in semiconductor nanostructures can be observed when impurities are inserted into the nanocrystal or by the simple fact that they exhibit large surface:volume ratios, and that the surface is not perfectly saturated, presenting dangling bonds. A larger degree of complexity appears when both impurity spins and surface magnetizations are considered. This leads to the possibility of an exchange interaction between the delocalized spins at the surface and the localized impurity states.

A prototype case for studying this system is Co-doped ZnO nanocrystals (Schoenhalz and Dalpian 2013). There are plenty of theoretical and experimental results for the bulk system, with strong controversies about its ground state. Co has seven d electrons, and when inserted into ZnO, these d levels will be placed at the band gap. The total magnetic moment per Co atom is $3\mu_B$, because the $e$ levels are lower in energy than the $t_2$ levels. Fig 4 shows the most important eigenvalues of three different nanocrystals, and a comparison to the non-doped nanocrystals. For bulk ZnO, as the band gap is reduced in ab initio calculations, the Co d levels might be incorrectly placed away from the bandgap. Nanocrystals, with larger bangaps, do not suffer from these problems.

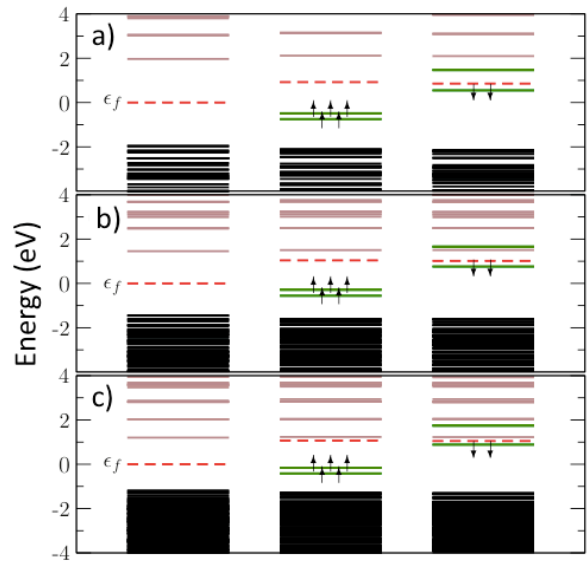

Fig. 4: Comparison of energy levels of a doped and undoped zinc blende NC. The set of eigenvalues in the left are for the pure, undoped nanocrystal, and the second and third sets represent the majority and minority spin channels, respectively. The undoped NC has (a) 35, (b) 87 and (c) 147 atoms of Zn and O. The doped NC had one of the Zn atoms (the most central) substitutedby a Co atom for each nanocrystal.

When two impurities are present inside the nanocrystal, the most stable magnetic coupling is antiferromagnetic, since no carriers are present. This leads to a controversy because experimental results report a ferromagnetic configuration for this kind of system. Our results indicate that this ferromagnetism can be understood by the existence of the surfaces, as reported before for pure ZnO nanocrystals.

To simulate the effect of the surfaces on doped nanocrystals, we have used a procedure similar to the one used for non-doped nanocrystals: we remove the surface saturation, insert an impurity inside the nanocrystal and optimize its structure. A very rich variety of spin configurations has been observed, depending on the size, structure, surfaces and number of dopants inside the nanocrystal. Besides the usual hybridization and crystals field splittings observed in bulk ZnO and on saturated ZnO nanocrystals, in non-saturated nanocrystals we also observe a strong interaction between the Co atom and the surface states, lowering the symmetry and tuning the occupation of the impurity atoms.

Figure 4 presents the radial spin density distribution for a Co-doped ZnO nanocrystal. In this case, there is only one impurity inside the nanocrystal. The curve in green is for the doped nanocrystal, with one Co atom near the center of the nanocrystal. For this case we observe a magnetization both at the center (Co atom) and at the surface of the nanocrystal. For the sake of comparison, we also show the magnetization for the non-doped nanocrystal. In this case, only the surface magnetization is observed.

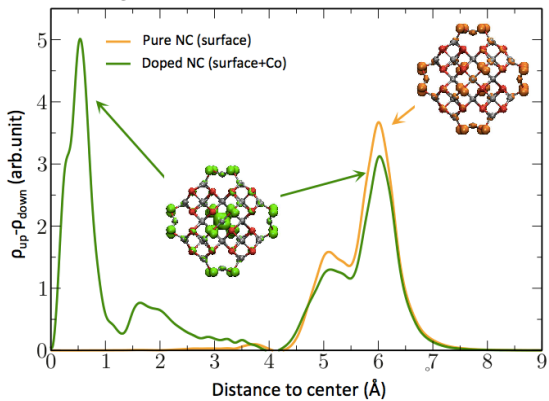

Fig. 4. Radial distribution of the spin density for ZnO nanocrystals. The green line indicates the case of the Co-doped nanocrystal, and the yellow line indicates the case for the bare nanocrystal. The insets show the real space representation of the spin density.

For the doped nanocrystal case, the removal of the surface saturation had two main effects. The first one was to induce a surface magnetization, as shown in Fig 4. The second effect is related to the magnetic coupling between two Co impurities inside the nanocrystal. For the saturated case, as there are no carriers in Co-doped ZnO, the most stable configuration between the spins of the Co atoms is antiparallel, i.e., antiferromagnetic. For unsaturated, reconstructed nanocrystals, the lowest energy configuration is when the spins of the Co atoms are parallel to each other, i.e., a ferromagnetic configuration. This occurs owing to the strong hybridization of the Co atoms with the surfaces, which might act as donors or acceptors changing the magnetic ground state. (Schoenhalz and Dalpian, 2013)

## 4. Conclusions and perspectives

The results presented in this chapter show the complex nature of spins in semiconductor nanocrystals. Owing to quantum confinement and surface effects, the behavior of these spins might be very different from their bulk counterparts. New phenomena are observed. Also, due to the reduced size of the nanocrystals, one has the possibility of inserting single impurities into them, leading to the possibility of single spin manipulation. This has potential applications ranging from quantum computing to spintronic devices.

The results presented in this chapter clearly show the importance of considering both quantum confinement and the surface structure for understanding the ground state properties of doped nanocrystals. As there are limited experimental methods that can precisely track the properties of these surfaces, theoretical methods become indispensable tools to understand their properties.

## Acknowledgements

This work was supported by Brazilian agencies FAPESP, CNPq and CAPES. GMD also thanks all the co-authors of the works reported in this chapter. Special thanks to Dr. A. L. Schoenhalz, whose PhD thesis focused on quantum confinement and surface effects in ZnO nanocrystals.